\documentclass[12pt]{article} 

\begin{document} 
\topmargin 0pt 
\oddsidemargin 0mm
\renewcommand{\thefootnote}{\fnsymbol{footnote}}
\begin{titlepage}
\vspace{5mm}

\begin{center}
{\Large \bf  A flat space-time relativistic explanation for the
  perihelion advance of Mercury } 
\end{center}
\vspace{6mm}
\begin{center}
{\large Harihar Behera$^a$\footnote{email: harihar@iopb.res.in},
    and P. C. Naik$^b$} \\
\vspace{5mm}
{\em
$^a$Patapur, P.O.-Endal, Jajpur-755023, Orissa, India\\ 
$^b$Department of Physics, D.D. College, Keonjhar-758001, Orissa, India}
\end{center}
\vspace{5mm}
\centerline{{\bf {Abstract}}}
\vspace{5mm}
Starting with the flat space-time relativistic versions of
Maxwell-Heaviside's  toy model vector theory of gravity and introducing 
the gravitational analogues for the electromagnetic Lienard-Wiechert potentials
together with the notion of a gravitational Thomas Precession; the observed
anomalous perihelion advance of Mercury's orbit is here explained as a
relativistic  effect in flat (Minkowski) space-time, unlike Einstein's
curved space-time relativistic explanation. In this new explanation for
the old paradoxical observation of Mercury's  perihelion shift, the
predicted value of the effect happens to coincide with Einstein's
predicted value in General relativity.\\


{\bf PACS}: 04.80Cc ; 96.30Dz.                       \\

{\bf Keywords} : {Gravitational Lienard-Wiechert potentials ,
  Gravitational Thomas Precession, Perihelion Advance of Mercury}.
\end{titlepage}
\section{Introduction}
The explanation of the anomalous perihelion shift of Mecury's orbit
represents one of the most famous classical tests of General 
Relativity (GR)- the relativistic gravitation theory in curved
space-time. This is the only one involving relativistic effects on
massive bodies - all other classical tests of GR being based on light
propagation effects only. The perihelion advance of Mercury had been an 
unsolved problem in celestial mechanics for over half a century, since 
the announcement by Leverrier in $1859$ \cite{1} that, after the
perturbing effects of the precession of the equinoxes on the
astronomical co-ordinate system had been subtracted, there remained
in the data an unexplained advance in the perihelion of Mercury. The
modern value of this discrepancy is $43$ arc-seconds per
century\cite{2}. A number of {\em ad hoc} proposals were made in an attempt 
to account for this excess, including , among others, to postulate
\cite{3,4,5} a magnetic-type component in gravity ( the gravitomagnetic
field) for the gravitational influence of the Sun on the motion of
planets ( the magnitude of this {\em  ad hoc} component could be adjusted so
as to account for the excess perihelion motion of Mercury), the
existence of a new planet Vulcan near the Sun, a ring of planetoids, a 
solar quadrupole moment and a deviation from the inverse-square law of 
gravitation, but none was successful. The great interest in this
paradoxical result vanished in 1915 when Einstein in his GR showed
that the excess advance of Mercury's perihelion could be explained as
a relativistic effect in curved space-time. In the GR the non-Newtonian 
``excess'' advance of perihelion of Mercury's orbit is explained by
using space curvature and Schwartzschild metric. Since the special
relativistic approach to the problem of perihelion advance in the
Kepler motion could not yield the observed effect (the existing
approaches yield only one sixth of Einstein's predicted value in GR, see for example \cite{6,7} ), the observation of this effect played a role of one of
the seemingly successful tests of the GR.  In this work we, by the way reporting the establishment\cite{8} of a compatibility of Newtonian gravity with
special relativity in a non-general-relativistic way, offer a flat
space-time relativistic explanation for the observed anomalous
perihelion advance of Mercury from a new angle. In this new approach
to the old and general relativistically solved problem we make use of
the notions  of gravitational Lienard-Wiechert potentials (GLWP) and the
 gravitational Thomas precession(GTP) which naturally follow from some
 extended special relativistic considerations. To  this end we devote
 Sec.$2$ to the notion of  gravitational Lienard-Wiechert potential
 (GLWP). In Sec.$3$ we show that the GLWP together with the GTP
 invoked in \cite{9} is sufficient to explain the observed perihelion
 advance of Mercury within the framework of a flat space-time
 relativistic toy model vector theory of gravity named as ``Maxwellian Gravity'' by the authors of \cite{8}.In Sec.$4$ we have some concluding remarks of interest for relativistic gravity.\\
\section{The  notion of GLWP }
The notion of gravitational Lienard-Wiechert potential (GLWP) is the
gravitational analogue of Lienard-Wiechert potential in classical
electromagnetism. So to invoke the GLWP one has to establish a
correspondence between the  gravitation theory and electromagnetic
theory. The close formal analogy between Newton's law of gravitation
and Coulomb's law of electricity led many authors, in the past and
also more recently , to investigate further similarities , such as the 
possibility that the motion of gravitational mass/charge could
generate the analogs of a magnetic field.The magnetic field is
produced by the motion of electric-charge,i.e. the electric current:
the motion of gravitational mass/charge would produce what is called
``gravitomagnetic'' field. It is to be noted that such a
correspondence had been proposed by J.C. Maxwell in 1865\cite{10}
and later further studied by Oliver  Heaviside\cite{11,12} and
O. Jefimenko\cite{13}.In the linearized versions of Einstein's
gravitational field equations, Maxwell-Lorentz-like equations for
gravity have been obtained by several authors, see for
examples\cite{3,14,15,16,17,18}. However from  a closer look at these
equations the reader can find that  the linearized theory of GR is not perfectly
isomorphic with electromagnetism which is commonly understood as a
limitation of Linearized GR\cite{17,18,19}. On the other hand there is
no such limitation in the linear relativistic gravity developed in
 \cite{8} under the name of ``Maxwellian Gravity''(MG).In its
formulation the relativistic nature of  gravity and the
source of gravity have been re-investigated by  re-designing  an often 
cited \cite{20,21} thought experiment \cite{22} involving the motion of two
point-like charged particles in Minkowski space-time.In the thought
experiment we have considered a system of two point-like charged particles having such amount of rest masses and charges that the force of electrostatic
repulsion balances that of gravitational attraction between the particles in
an inertial frame of reference \(K'\) in which both the particles are at
rest under equilibrium condition. Then we investigated in\cite{8} the
condition of equilibrium of the said particle system in \(K'\)-frame
as well as in another inertial frame $K$ moving with uniform velocity
with respect to the \(K'\)-frame of reference. From the requirement of
the frame-independence of the equilibrium conditions, we not only obtained a
Lorentz-force law for gravitational interaction between the moving masses as
expected \cite{22} but also unexpectedly found the Lorentz-invariant rest
masses of the interacting particles as representing their
gravitational masses (or gravitational charges) in complete analogy
with the Lorentz force law and the invariant electric charges of the
classical  electromagnetic theory. These findings are in conformity with
Poincar\'{e}'s\cite{23} remark that {\em if
equilibrium is to be a frame-independent condition, it is necessary for all
forces of non-electromagnetic origin to have precisely the same
transformation law as that of the Lorentz-force.} Having
recognized these findings we have obtained four Faraday-Maxwell-type toy-model linear equations of gravity describing what we call
``Maxwellian Gravity'' following the known procedures of the
electromagnetic theory. The equations have a surprisingly rich and detailed
correspondence with Faraday-Maxwell's field equations of the electro-magnetic
theory. The field equations can be written in the following Faraday-Maxwellian-form:\\
\begin{equation}
\vec{\nabla}\cdot{\vec{E}_{g}} = -4\pi G \rho_{0} =
-{\rho_{0}}/{\epsilon_{0g}}
\; \;\;\;\;\;\; \; \; {where \;\;\;\;  \epsilon_{0g}={1}/4\pi G}
\end{equation}
\begin{equation}
\vec{\nabla} \times \vec{B}_g = - \mu_{0g} \vec{j}_0 + (1/c^2)
({\partial \vec{E}_g }/{\partial t}) ,\;\;\;\;{where\;\;\;\; \mu_{0 g}
= {4{\pi}G}/{c^{2}}}
\end{equation}
\begin{equation}
\vec{\nabla} \cdot \vec{B}_{g} = 0
\end{equation}
\begin{equation}
\vec{\nabla} \times \vec{E}_g = - \partial \vec{B}_g / \partial t
\end{equation}
Where $\rho_{0}$ = rest mass (or proper mass)  density; $\vec{j_{0}}$
= rest mass current density; $G$ is Newton's universal gravitational
constant; $ c $ is the speed of light in empty space; the gravito-electric and gravito-magnetic
fields $\vec{E_{g}}$ and $\vec{B_{g}}$  respectively are defined by
the gravitational Lorentz force on a test particle of rest mass $
m_{0} $ moving with uniform velocity $ \vec{u} $ as
\begin{equation}
\frac{d}{dt}[m_{0}\vec{u}/(1-{u^2}/{c^2})^{1/2}] = m_{0}[\vec{E_{g}}+\vec{u}\times\vec{B_{g}}]
\end{equation}
where the symbols have their respective meanings in correspondence with
the Lorentz force law in its relativistic form.It is to be noted that
these set of gravitational Maxwell-Lorentz equations  coincide with
those speculated by Maxwell\cite{10}, Heaviside\cite {11,12} and
discussed also by Peng\cite{17} in the weak field and slow motion limit of
Einstein's field equations. However the treatment made in \cite{8} 
suggests the  validity of these equations for relativistic speeds as well. 
In covariant formulation, introducing the space-time four vector
$x_{\mu} = (x,y,z,ict)$, proper
mass current density four vector $j_{\mu}=(j_{0x},
j_{0y},j_{0z},ic\rho_{0})$  and the second-rank antisymmetric
gravitational field strength tensor
\begin{eqnarray}
F_{\mu\nu} = \pmatrix {
0 & B_{gz} & -B_{gy} & -iE_{gx}/c\cr
-B_{gz} & 0 & B_{gx} & -iE_{gy}/c \cr
B_{gy} & -B_{gx} & 0 & -E_{gz}/c \cr
iE_{gx}/c & iE_{gy}/c & iE_{gz}/c & 0}
\end{eqnarray}
The field equations (1-4) can now be represented by the following two
equations:
\begin{equation}
\sum_{\nu}{\partial F_{\mu \nu}}/{\partial x_{\nu}} = -\mu_{0g}{\vec j_{\mu}}, \; \;  {\rm where} \; \; \mu_{0g} = 4{\pi}G/{c^{2}}
\end{equation}
\begin{equation}
\partial F_{\mu \nu}/ \partial x_{\lambda} +\partial F_{\nu \lambda}/ {\partial x_{\mu}} + \partial F_{\lambda \mu}/ {\partial x_{\nu}} = 0
\end{equation}
while the gravitational Lorentz force law (5) assumes the form :
\begin{eqnarray}
c^{2}(d^2 x_{\mu}/ds^2) = F_{\mu \nu}({dx^{\nu}}/ds)
\end{eqnarray}
The absence of the rest mass of the test particle in its co-variant equation
of motion (9) in the external gravitational field $ F_{\mu\nu} $
describes clearly the universal nature of gravitational interaction in
conformity with Galileo's empirical law of universality of free fall
(UFF) in a uniform gravitational field in the relativistic case as
well. The law of UFF is one of the most fascinating character of
gravity. For the Newtonian theory, the universality of free fall is of course a
consequence of the equality of inertial mass and gravitational
mass. Being fascinated by the mystery of the UFF observed in Galileo-Newtonian
physics,  Einstein in his development of the
GR postulated that it holds generally, in particular also for large
velocities (the relativistic case) and strong fields and for any form
of mass-energy (i.e. he postulated the equality of inertial mass and
gravitational mass for the UFF to hold generally). It is to be
carefully noted that  the general validity of the UFF is now a natural
consequence of Maxwellian Gravity without the requirement of an
Einsteinian  postulate on the equality of inertial and gravitational
masses. This is one of the interesting theoretical revelations of MG
exacting the real UFF. The origins of the UFF may be traced to the
Lorentz-invariant nature of the gravitational mass-charge (i.e. the rest mass
or any form of Lorentz-invariant mass-energy) and the laws of physics
as dictated by special relativity.

The fields $ {\vec E_{g}} $   and ${ \vec B_{g}} $ of MG
are derivable from potential functions
\begin{equation}\vec B_{g} = \nabla\times{\vec A_{g}},\>\>\>\>\>\>\>\>  \vec E_{g} = -\nabla \cdot \Phi_{g} - {\partial{\vec A_{g}}/{\partial t}}
\end{equation}
where ${ \Phi_{g}} $ and ${ \vec A_{g}} $ represents respectively the gravitational scalar and vector potential of MG.These
potentials satisfy the inhomogeneous wave equations :
\begin{equation}\nabla^{2}\cdot\Phi_{g}\, - \,\frac{1}{{c}^{2}}\cdot\frac{\partial^{2}\Phi_{g}}{\partial t^{2}}\,=\,4\pi\,G\, \rho_ {0}\,=\,\rho_{0}/\epsilon_ {0\,g}  
\end{equation}
\begin{equation}\nabla^ {2}\cdot\vec A_{g}\, -\,
  \frac{1}{c^{2}}\cdot\frac{\partial^{2}\vec A_{g}}{\partial
    t^{2}}\,=\,\frac{4\pi\,G}{c^{2}}\vec j_{0}\,= \,\mu_{0\,g}\vec j_{0} 
\end{equation}
if the gravitational Lorenz\cite{24} gauge condition
\begin{equation}\vec{\nabla}\cdot\vec
  A_{g}\,+\,\frac{1}{c^{2}}\frac{\partial\Phi_{g}}{\partial{t}}\,=\,0
\end{equation}
is imposed. These will determine the generation of gravitational waves 
by prescribed gravitational charge and current
distributions. Particular solutions (in vacuum) are 
\begin{equation}\Phi_{g}\,(\,\vec r\,,t\,)\,=\,-\,G\,\int\,{\frac{\rho_{0}(\,\vec r^{\prime}\,,\,t^{\prime}\,)}{\,|\vec r\,-\,\vec r^{\prime}\,|}dv^{\prime}}
\end{equation} 
\begin{equation}\vec A_{g}\,(\,\vec
  r\,,t\,)\,=\,-\,\frac{G}{c^{2}}\,\int\,{\frac{\vec j_{0}(\,\vec r^{\prime}\,,\,t^{\prime}\,)}{\,|\vec r\,-\,\vec r^{\prime}\,|}dv^{\prime}}
\end{equation}
 where $\,t^{\prime}\,= \,t\,-\,{|\,\vec r\,-\,\vec
   r^{\prime}\,|}/c\,$  is the retarded time.These are called the
 retarded potentials. Thus we saw that retardation in gravity is
 possible in Minkowski space-time in the same procedure  as we adopt
 in electrodynamics.This result seems to conflict with the view
 \cite{25} that Newtonian gravity is entirely static, retardation is
 not possible until the correction due  to deviations from Minkowski
 space is considered.                                         
Now in analogy with the electromagnetic case we have here the
Gravitational Lienard-Wiechert potentials for a point particle of rest 
mass $m_{0}$ moving with velocity $\,\vec v\,$ as
\begin{equation}\Phi_{g}\,=\,-\,\frac{G\,m_{0}}{\,r\,(\,1\,-\,\frac{v^{2}\sin^{2}\theta}{c^{2}})^{1/2}}
\end{equation}
\begin{equation}\vec A_{g}\,=\,-\,\frac{G\,m_{0}\,\vec v}{\,c^{2}\,r\,(\,1\,-\,\frac{v^{2}\sin^{2}\theta}{c^{2}})^{1/2}}
\end{equation}
where $\,r\,$ is the magnitude of the instantaneous  position vector
$\vec r$ of the field  point from the position of the particle and $\theta$ is
the angle between $\,\vec r\,$ and $\,\vec v\,$ at the instant of time. Thus for a planet with rest mass $\,m_{0}\,$ having relative velocity
$\,\vec v\,$ with respect to the the Sun (with rest mass
$\,M_{\odot}\,$) the instantaneous gravitational Lienard-Wiechert (LW) potential energy is given by
\begin{equation}U_{gLW}\,=\,-\,\frac{G\,M_{\odot}\,m_{0}}{\,r\,(\,1\,-\,\frac{v^{2}\sin^{2}\theta}{c^{2}})^{1/2}}
\end{equation}
Considering  the angular momentum of the planet as
$\,L\,=\,m_{0}rv\sin\theta\,$ Eq.(18) can be re-written as 
\begin{equation}U_{gLW}\,=\,-\,\frac{G\,M_{\odot}\,m_{0}}{\,r\,(\,1\,-\,\frac{\,L^{2}}{m_{0}^{2}\,c^{2}\,r^{2}})^{1/2}}
\end{equation}
which in the first approximation can be reduced to
\begin{equation}U_{gLW}\,=\,-\,\frac{G\,M_{\odot}\,m_{0}}{r}\,-\,\frac{G\,M_{\odot}\,L^{2}}{2\,m_{0}\,c^{2}\,r^{3}}\,=\,-\,\frac{k}{r}\,-\,\frac{h_{0}}{r^{3}}
\end{equation}
where 
\begin{equation}\,k\,=\,GM_{\odot}m_{0}\,\,\,\mbox{ and} \,\,\,h_{0}\,=\,\frac{G\,M_{\odot}\,L^{2}}{2\,m_{0}\,c^{2}\,}\,
\end{equation}
Thus we saw that the consideration of the GLWP introduced a $\,1/r^{3}\,$
potential into the Kepler problem. Its effect on the planetary motion
will be considered in the following section together with the GTP effect.
\section{The GTP, GLWP and The perihelion advance }
 The Thomas precession\cite{26,27,7} is purely kinematical in origin
\cite{27}. If a component of acceleration $ (\vec a)$ exists
perpendicular to the velocity $ \vec v $, for whatever reason, then
there is a Thomas Precession, independent of other effects
\cite{27}. When the acceleration is caused by a gravitational force
field, the corresponding Thomas Precession is reasonably referred to
as the Gravitational Thomas Precession (GTP). Given the physics
involved in the Thomas Precession, the possibility of the existence of 
the GTP in planetary motion can not be ruled out in
principle.
The Thomas Precession frequency $ \vec\omega_{T} $ in the
non-relativistic
 limit (i.e., when $ v << c $) is given by \cite{7,27} 
\begin{equation}
\vec\omega_{T}\,=\,\frac{1}{2c^{2}}(\vec a \times \vec v ), 
\end{equation} 
where the symbols have there usual meanings. For a planet (say Mercury)
 moving around the Sun, the acceleration $ \vec a $ is predominately
 caused 
 by the Newtonian gravitational field of the Sun,viz., 
\begin{equation} 
\vec a\,=\,-\,\frac{GM_{\odot}}{r^{3}}\,\vec r\,\,, 
\end{equation} 
where the symbols have their usual meanings. Thus, from Eqs.$(22)$ and
$(23)$ we
 get the GTP frequency of the planet in question as 
\begin{equation} 
\vec \omega_{gT}\,=\,-\,\frac{GM_{\odot}}{2c^{2}r^{3}}\,(\vec r \times
\vec v )\,\,, 
\end{equation} 
where $ \vec v $ is the velocity of the planet. In terms of the angular
momentum of the planet  $\,\vec L\,=\,m_{0}(\vec r\times\vec v)\,$,
Eq.$(24)$ can be re-written as
\begin{equation} 
\vec \omega_{gT}\,=\,-\,\frac{GM_{\odot}}{2m_{0}\,c^{2}r^{3}}\,\vec L\,\,, 
\end{equation}
If, as Thomas first pointed out, that coordinate system rotates, then
the total time rate of change of the  angular momentum $\,\vec
J\,$\footnote{For Thomas $\vec J\,=\,\vec S\,$, the spin angular
  momentum; but here we consider a more general term $\vec J\,=\,\vec
  L\,+\,\vec S\,$, $\vec J\,$ representing the total ( orbital\,+\,spin\,) angular momentum of the particle under 
  consideration. } or more
generally, any vector $\,\vec A\,$ is given by the well known result
\cite{7,27},
\begin{equation} \left(\frac{d\vec A}{dt}\right)_{\mbox{nonrot}}\,\,\,=\,\left (\frac{d\vec A}{dt}\right)_{\mbox{rest\,\,frame}}\,\,\,\,+\,\,\vec\omega_{T}\,\times\,\vec A
\end{equation}
where $\,\vec\omega_{T}\,$ is the angular velocity of rotation found
by Thomas. When applied to the total angular momentum $\,\vec J\,$,
Eq.$(26)$ gives an equation of motion:
\begin{equation} \left(\frac{d\vec J}{dt}\right)_{\mbox{non-rot}}\,\,\,=\,\left(\frac{d\vec J}{dt}\right)_{\mbox{rest\,\,frame}}\,\,\,\,+\,\,\vec\omega_{T}\,\times\,\vec J
\end{equation}
The corresponding energy of interaction is
\begin{equation} U\,=\,U_{0}\,+\,\vec J\,\cdot\,\vec\omega_{T}\,=\,U_{0}\,+\,\vec L\,\cdot\,\vec\omega_{T}\,+\,\vec S\,\cdot \vec\omega_{T}
\end{equation}
where $\,U_{0}\,$ is the energy corresponding to the coupling of
$\,\vec J\,$ to the external fields - say the Coulomb field in atomic
case ,nuclear field in nuclear case and the Newtonian gravitational
field in the planetary case. The origin of the Thomas precessional
frequency $\,\vec\omega_{T}\,$ is the acceleration experienced by the
particle as it moves under the action of external forces\cite{27}. Since 
the nature of the external forces is not specified, the result
obtained in Eq.(28) is valid for all type of force fields which cause
accelerations of whatever nature. When applied to the gravitodynamic
problems in solar system where the acceleration of a planet with
respect to the Sun is predominately  caused
by a force arising out of the Newtonian scalar potential, Eq.(28) takes the form
\begin{equation} U_{g}\,=\,U_{0g}\,+\,\vec J\,\cdot\,\vec\omega_{T}\,=\,U_{0g}\,+\,\vec L\,\cdot\,\vec\omega_{gT}\,+\,\vec S\,\cdot \vec\omega_{gT}
\end{equation}
where $\,U_{0g}\,$ is the Newtonian potential energy of the planet
under consideration and $\,\vec\omega_{gT}\,$ is given by Eq.(25).We
then have
\begin{equation} U_{g}\,=\,-\,\frac{k}{r}\,-\,\frac{h_{1}}{r^{3}}\,-\,\frac{h_{2}}{r^{3}} 
\end{equation}
where $\,k\,=\,GM_{\odot}m_{0}\,$ and  
\begin{equation}h_{1}\,=\,\frac{GM_\odot L^{2}}{2m_{0}\,c^{2}} 
\end{equation}
\begin{equation}h_{2}\,=\,\frac{GM_\odot}{2m_{0}\,c^{2}}(\vec L\cdot\vec S). 
\end{equation}
Thus we see the gravitational Thomas precession in the
non-relativistic limit introduced two
potentials of the form $\,1/r^{3}\,$ into the classical Kepler
problem. If in place of the $\,U_{0g}\,$ in Eq.(29) we take the
gravitational Lienard-Wiechert (LW) potential energy $\,U_{gLW}\,$
given by Eq.(20),  we would then  have the following
equation in place of Eq.(30) :
\begin{equation} U_{g}\,=\,-\,\frac{k}{r}\,-\,\frac{h_{0}}{r^{3}}\,-\,\frac{h_{1}}{r^{3}}\,-\,\frac{h_{2}}{r^{3}}\,=\,\frac{k}{r}\,-\,\frac{h_{MG}}{r^{3}}
\end{equation}
where
\begin{equation} h_{MG}\,=\,h_{0}\,+\,h_{1}\,+\,h_{2}\,=\,\frac{GM_\odot L^{2}}{m_{0}\,c^{2}}\left[\,1\,+\,\frac{(\,\vec L\,\cdot\,\vec S\,)}{2\,L^{2}}\,\right]. 
\end{equation}
What effect will result from the introduction of the potential(s) of the
form $\,1/r^{3}\,$ into the Kepler problem ? It is  shown in \cite{7}
that if a potential with $\,1/r^{3}\,$  form is added to a central
force perturbation of the bound Kepler problem, the orbit in the bound
problem is an ellipse in a rotating coordinate system. In effect the
ellipse rotates, and the periapsis appears to precess. If the perturbation Hamiltonian is
\begin{equation}\bigtriangleup H\,=\,-\,\frac{h}{r^{3}},\,\,\,\,\,\,\,\,\,(\mbox{\,h\,=\,some constant\,})\,
\end{equation}
then it predicts \cite{7} a precession of the perihelion of a planet arising
out of the perturbation Hamiltonian (of the form as in Eq.(35)) at an
average  rate of
\begin{equation} 
\dot{\tilde{\bf\omega}}\,=\,\frac{6 \pi\,k\,m_{0}^{2}h }{\,\tau\,L^{4}}
\end{equation}
where $\,k\,=\,GM_{\odot} m_{0} $ and $\,\tau\,$ is the classical period of
revolution of the planet around the sun. It is worth-noting from \cite{7,28}  that the so-called Schwarzschild spherically symmetric solution of the Einstein field equations corresponds to an additional Hamiltonian in the Kepler problem of the form of Eq.(35) with
\begin{equation} h\,=\,h_{E}\,=\,\frac{GM_{\odot}L^{2}}{m_{0}\,c^{2}}
\end{equation}
so that Eq.(36) becomes 
\begin{equation}
\dot{\tilde{\bf\omega}}_{E}\,=\,\frac{6\,\pi\,k^{2}\,}{\,\tau\,L^{2}\,c^{2}}\,=\, \frac{6\,\pi\,GM_{\odot}}{\,\tau\,c^{2}\,a\,(\,1\,-\,e^{2}\,)}
\end{equation}
where we have used the relation
$\,L^{2}\,=\,GM_{\odot}\,m_{0}^{2}\,a\,(\,1\,-\,e^{2})\,$. Eq.(38)
represents Einstein's expression for the anomalous perihelion advance
of a planet's orbit. Likewise the contributions to the perihelion
advance arising out of the GLWP and GTP in the framework of Maxwellian 
Gravity can be estimated by
taking the $\,h\,$ in Eq.(36) as
\begin{equation}\,h\,=\,h_{MG}\,=\,\frac{GM_\odot L^{2}}{m_{0}\,c^{2}}\left[\,1\,+\,\frac{(\,\vec L\,\cdot\,\vec S\,)}{2\,L^{2}}\,\right]\,=\,h_{E}\left[\,1\,+\,\frac{(\,\vec L\,\cdot\,\vec S\,)}{2\,L^{2}}\,\right].
\end{equation}
Then Maxwellian Gravity can predict the relativistic perihelion
advance of a planet at  
\begin{equation}
\dot{\tilde{\bf\omega}}_{MG}\,=\,\dot{\tilde{\bf\omega}}_{E}\,\left[\,1\,+\,\frac{(\,\vec L\,\cdot\,\vec S\,)}{2L^{2}}\,\right].
\end{equation}
For Mercury, the value of
$\,\dot{\tilde{\bf\omega}}_{E}\,=\,42\cdot98\,$ arc-seconds/century - a
well known data \cite{2,7,29,30}. Hence the relativistic 
perihelion advance of Mercury's orbit in the flat space-time
Maxwellian Gravity could be predicted at
\begin{equation}
\dot{\tilde{\bf\omega}}_{MG}\,=\,42\cdot98\,\left[\,1\,+\,\frac{(\,\vec
    L\,\cdot\,\vec S\,)}{2L^{2}}\,\right] \mbox{arc-seconds/century}
\end{equation}
The additional $ L-S $ term viz., $\,\frac{(\,\vec L\,\cdot\,\vec
  S\,)}{2L^{2}}\,$, that appears in Eq.(41) has a numerical value of
the order of $\,10^{-10}\,$ when the physical and orbital parameters
of Mercury are used and is therefore  utterly negligible. So by neglecting this $ L-S $ term we get                                                                          \begin{equation}
\dot{\tilde{\bf\omega}}_{MG}\,=\,\dot{\tilde{\bf\omega}}_{E}\,=\,\,42\cdot98\,\,\,\mbox{arc-seconds/century \,\,for\,\,Mercury}.
\end{equation}                                                                 \section{Concluding Remarks}
     In this work we saw the possibility of explaining the observed
     anomalous advance of the perihelion of Mercury's orbit in flat
     space-time relativistic gravity. This new approach to the old
     gravitodynamic problem  of perihelion advance may serve as a test
     of the validity of special relativity in the domain of
     gravitation. Again this also implies a test of the physics of
     Thomas Precession in gravitational phenomena. It is to be noted
     that forces and accelerations (of whatever origin) are well
     within the scope of special relativity (SR) because the SR in its 
     entirity no where forbids one to study the force of gravity
     within its versatile scope. In this connection we would like to
     quote an important observation made by Denisov and
     Logunov\cite {31}:\begin{quote}{\em`` ...,it must be noted that the literature not infrequently contains statements claiming that the special theory of relativity deals with the description of phenomena in the inertial reference frames, while the description of phenomena in non-inertial reference frames is the prerogative  of the GTR.\\
These statements are wrong. ...... . Because of this, it is quite
conceivable to describe the physical phenomena either by the special
theory of  relativity or within non-inertial reference frames. This
point was trasparently clear to Fock [13]. ''}\end {quote}
By the way we remark that we are not proposing a new theory of gravity as MG. In our work on MG we only  investigated some unexplored aspects of relativistic gravity in flat
space-time and eleveted the status of Maxwell-Heaviside's gravity to that of a test theory for testing the foundations of both special and general
relativity. It is to be carefully noted that MG is now a natural
outcome of some well established principles,theories and methods of  
study in physics. Therefore the predictions of the MG may not be
totally false. The theory might be working somewhere in some domain of
physics yet unexplored. So we have to explore the situations where and 
when the MG was/is/might be operating in the evolution of the physical 
world. The authors make an appeal to the redears not to consider the MG 
as an alternative theory of gravity to the GR, because MG has to be made
compatible with  many other experinental data or observational results 
for its  elevation to that status. So we now prefer MG to be treated as a
toy model vector theory of gravity in flat space-time.

\textbf{Acknowledgments}
The authors wish to thank Prof. Lewis Ryder, Phys. Laboratory, University of
Kent, Canterbury, UK; Prof. B. Mashhoon, Dept. of Phys. and Astronomy,
Univ. of Missouri-Columbia, Columbia, Missouri, USA and
Prof. M. Gasperini, Dept. of Phys., Univ. of Bari, Bari, Italy for
their careful reading of the basic paper \cite{8} and for their kindly
e-mailed valuable critical remarks and constructive criticisms which
we found encouraging and of some use for this work. The first author
would like to thank Prof. N. Barik and Prof. L. P. Singh, both of  Dept. of
Phys., Utkal University, Vani Vihar, Bhubaneswar, Orissa, for their kind
co-operation in some thought provoking and insightful  discussions on the subject with them. The authors also acknowledge  the help received from the Institute of Physics, Bhubaneswar for using  its library and computer Centre for this work.
                
\end{document}